\documentclass[iop]{emulateapj}

\usepackage{graphicx}
\usepackage[suffix=]{epstopdf}
\usepackage[export]{adjustbox}
\usepackage{natbib}
\usepackage{amsmath}
\usepackage{url}
\usepackage{xspace}

\newcommand{\Kepler}{\textsl{Kepler}\xspace}

\begin{document}
\title{Detecting Differential Rotation and Starspot Evolution on the\\ M dwarf GJ 1243 with \Kepler}

\shorttitle{Starspot Evolution on GJ 1243}
\shortauthors{Davenport et al.}

\author{
	James R. A. Davenport\altaffilmark{1,2},
	Leslie Hebb\altaffilmark{3}, 
	Suzanne L. Hawley\altaffilmark{2}
	}

\altaffiltext{1}{Corresponding author: jrad@astro.washington.edu}
\altaffiltext{2}{Department of Astronomy, University of Washington, Box 351580, Seattle, WA 98195}
\altaffiltext{3}{Department of Physics, Hobart and William Smith Colleges, Geneva, NY, 14456}
 
\begin{abstract}
We present an analysis of the starspots on the active M4 dwarf GJ 1243, using four years of time series photometry from \Kepler. A rapid $P = 0.592596\pm0.00021$ day rotation period is measured due to the $\sim$2.2\% starspot-induced flux modulations in the light curve. We first use a light curve modeling approach, using a Monte Carlo Markov Chain sampler to solve for the longitudes and radii of the two spots within 5-day windows of data. Within each window of time the starspots are assumed to be unchanging. 
Only a weak constraint on the starspot latitudes can be implied from our modeling. 
The primary spot is found to be very stable over many years. 
A secondary spot feature is present in three portions of the light curve, decays on 100-500 day timescales, and moves in longitude over time. We interpret this longitude shearing as the signature of differential rotation. Using our models we measure an average shear between the starspots of 0.0047 rad day$^{-1}$, which corresponds to a differential rotation rate of $\Delta\Omega = 0.012 \pm 0.002$ rad day$^{-1}$. 
We also fit this starspot phase evolution using a series of bivariate Gaussian functions, which provides a consistent shear measurement. This is among the slowest differential rotation shear measurements yet measured for a star in this temperature regime, and provides an important constraint for dynamo models of low mass stars.
\end{abstract}

\section{Introduction}

For low-mass, fully convective stars, the nature of the magnetic dynamo and the role of differential rotation is not so clear. Some radial and surface differential rotation is expected to exist, due to the combination of rotation and convection. However, despite the deep convective zones of M dwarfs, their long convective turnover timescales result in a lower amplitude of surface differential rotation or shear \citep{kuker_rudiger2008,kitchatinov2011}, (the $\Omega$ effect). Since these stars are nearly or fully convective from surface to core, and therefore lack a ``tachocline'' interface region in which to store toroidal magnetic field, the dynamo mechanism must be fundamentally different than the popular $\alpha \Omega$ dynamo model for the Sun \citep[e.g.][]{parker1955,schrijver_zwaan2000}. Instead, this convectively driven process is known as an $\alpha^2$ dynamo.  For rapidly rotating M dwarfs, the magnetic field strength is expected to be increased, which suppresses differential rotation and forces nearly solid-body rotation \citep{browning2008}. Without strong radial or surface differential rotation to organize the global magnetic field, activity cycles may not be present, and the surface magnetic topology is predicted by some models to be highly non-axisymmetric and multipolar \citep[e.g.][]{chabrier2006}.

However, observations of many low mass stars reveal highly organized, strongly poloidal magnetic fields \citep[e.g.][]{morin2008b} and prominent long-lived starspot features \citep[e.g.][]{barnes2005a}. Some rapidly rotating low-mass stars show indications of polar ``starspot caps'' possibly due to this large scale dipolar field \citep{donati1997,morin2008a}, while others do not \citep{barnes2004,morin2010}. 
Though differential rotation is expected to play a lesser role for these rapidly rotating low-mass stars, even small amounts of differential rotation may help to organize the chaotic, $\alpha^2$ driven magnetic fields in to a coherent, axisymmetric field \citep{kitchatinov2011}, capable of producing very long-lived polar spot features. Therefore, given the wide variety of observed surface magnetic topologies, and the complex inter-dependence of rotation, differential rotation, and the magnetic field, measuring differential rotation rates for low mass stars is a high priority for constraining dynamo theory.

Rotation can now be measured with relative ease for many stars, for example using spectral line broadening that produces v~sin~i measurements, or periodic flux modulations due to starspots in precision space-based time series photometry. Data from the \Kepler mission \citep{borucki2010} has revolutionized the study of stellar rotation using starspot modulations, with tens of thousands of stars having measuring rotation periods \citep{mcquillan2014,reinhold2013}, and has revealed starspot properties for stars ranging from solar mass \citep{bonomo2012} to brown dwarfs \citep{gizis2013}.

Differential rotation, however, is notoriously difficult to detect for stars. 
Spectral techniques can trace active regions at different latitudes for stars with lower activity levels such as the Sun \citep{bertello2012}. Detecting differential rotation via Zeeman Doppler Imaging \citep[ZDI;][]{semel1989,donati_brown1997} requires comparing complex surface magnetic reconstructions or maps between subsequent visits. 
Photometric surveys may be able to produce differential rotation rates for an ensemble of active stars \citep[][]{reinhold2013a}.
A recent blind survey of competing techniques for detecting rotation and differential rotation from model photometry showed excellent agreement in recovering rotation periods from active stars. However,  a complex degeneracy was found between differential rotation rate, starspot lifetimes, and the number of starspots present, and little agreement between competing methods and the model light curves \citep{aigrain2015}. 
In general, methods for detecting differential rotation in photometry follow one of two approaches:
1) Fourier methods, which measure the broadening or splitting of peaks in the power spectrum, auto-correlation function, or periodogram, or equivalently by decomposition of the light curve using sine functions \citep{reinhold2013}. These methods utilize the entire light curve at once, and are efficient for analyzing large volumes of data from many stars, but may suffer more from the degeneracies mentioned above. 
2) Tracking specific starspot features either via light curve inversion \citep{roettenbacher2013}, or light curve modeling for individual starspots \citep{frasca2011}. These methods are more computationally expensive, but their results seem robust for rapidly rotating stars with long-lived spots.

In this paper we venture into a relatively new region of starspot evolution parameter space, detecting very gradual differential rotation and spot decay for a rapidly rotating M dwarf. The fast time cadence and continuous monitoring provided by \Kepler, along with a short stellar rotation period, allow us to trace small changes in starspot phase and amplitude over long periods of time. 
In Section \ref{sect:period} we describe our target, the active M dwarf GJ 1243, and the previous investigations of this low-mass star with \Kepler. 
Our detailed light curve modeling is presented in Section \ref{sect:mcmc}.  
We trace small changes in starspot phase over four years, and interpret this as a signature of differential rotation in Section \ref{sect:trace}. 
A simpler approach to detect this slow differential rotation by modeling the phase evolution with Gaussians is given in Section \ref{sect:gauss}. 
We place the differential rotation signal from GJ 1243 in the context of other cool stars, and compare the \Kepler photometric results with older ground-based data in Section \ref{sect:discussion}.
Finally, in Section \ref{sect:summary} we provide a summary of our results, and discuss the great potential for understanding starspots and the stellar dynamo still to be realized from the unique photometric \Kepler database and future missions.

\begin{figure*}[]
\centering
\includegraphics[width=7in]{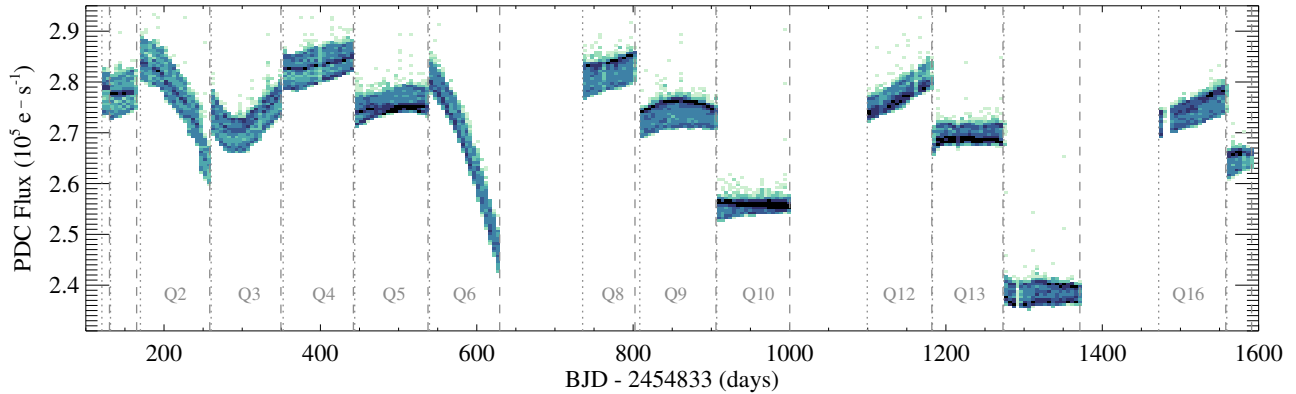}
\caption{The long cadence PDC MAP light curve for GJ 1243. Pixel shade (light to dark) indicates the density of epochs. Breaks in the light curve due to quarterly spacecraft rolls are indicated (grey dashed lines).}
\label{lc}
\end{figure*}

\section{GJ 1243}
The target of our study is the nearby mid M dwarf, GJ 1243 (Kepler ID \# 09726699). 
This star has a short rotation period of 0.5926 days that has been noted in previous studies of \Kepler light curves \citep[][]{savanov2011,mcquillan2013}.
The spectral type has been measured as M4 \citep{hawley2014}, placing GJ 1243 near the fully-convective boundary where stars are expected to remain magnetically active for many Gyr \citep{nlds, west2008}.
Using the parallax distance of 11.9 pc from \citet{lepine2005}, the apparent $K$-band magnitude from \citet{ucac4}, and the M$_{K}$-mass relation from \citet{delfosse2000}, we estimate a mass for GJ 1243 of 0.24~${M_{\odot}}$. The convective turnover timescale for stars in this mass range (assuming M=0.235$M_\odot$) from \citet{kiraga2007} is quite slow at $\tau_{conv} \approx 70$ days. Comparing this timescale to the rotation period, we find GJ 1243 has a very low  Rossby number of $R_o=P_{rot}/\tau_{conv}\approx0.008$. 
Lucky imaging of GJ 1243, as well as ground based spectroscopy, have shown no indications of a binary companion (Wisniewski 2015 in preparation). 
In addition, GJ 1243 has been the subject of detailed flare activity studies with \Kepler data, producing the largest catalog of M dwarf flares ever observed for a single star \citep{hawley2014,davenport2014b}.
In this section we describe our treatment of the \Kepler data for this active M dwarf, removing systematic trends from the light curve, and detecting a periodic signal.

\subsection{Kepler Long Cadence Data}
The \Kepler light curve for GJ 1243 contains dramatic stellar variability in the form of flares and starspots. \citet{ramsay2013} have examined the flare energy distribution using one quarter (Q6) of data from \Kepler. \citet{davenport2014b} and \citet{hawley2014} used 11 months of \Kepler short cadence (1-minute) data for GJ 1243, over 300 days worth in total, to robustly measure the flare rate and develop a statistical understanding of the flare morphology from this very active dwarf. For these flare studies the starspot signature had been treated as a noise source to be smoothed out.

In the present investigation, we utilized all available long cadence (30-minute) \Kepler data for GJ 1243 to study the evolution of the starspots while minimizing the impact of small amplitude flares. GJ 1243 was observed in 14 separate quarters of \Kepler data (Q0--Q6, Q8--Q10, Q12--Q14, and Q16--Q17), spanning over four years of observation (MJD 54953.04 through 56423.50). We used the most recent reduction of the \Kepler data available, including the ``PDC-MAP'' Bayesian de-trending analysis from \citep{smith2012}.
The entire 4-year catalog PDC-MAP light curve for GJ 1243 is shown in Figure \ref{lc}. Data from Q7, Q11, and Q15 was not available due to the failure of CCD Module 3 in 2010, which GJ 1243 resided on for one quarter of the year.

In Figure \ref{lc}, large discontinuities in the flux are apparent between quarters, as well as systematic trends in the mean flux within quarters. These long timescale variations are systemic to \Kepler data, due to spacecraft drift and calibration limitations, and are not astrophysical. 
For every quarter, we fit and subtracted low order (linear or quadratic) polynomials from the data to remove these systematic errors and discontinuities. 
Because the stellar rotation period is so short, and each quarter contains on average $\sim$150 rotations, these polynomial fits do not affect the starspot signal on the timescales we are interested in.

Large amplitude flares were also present in our data, visible as positive flux excursions throughout the light curve in Figure \ref{lc}. While the short cadence \Kepler data for GJ 1243 is a treasure trove for flare studies \citep[e.g.][]{davenport2014b}, only the largest energy flares are visible in the 30-minute data \citep[see][]{walkowicz2011}. To remove the flares from our analysis, we smoothed the light curve with a 12-hour ``boxcar'' filter, and then discarded epochs with fluxes that deviated by more than 0.3\% from the smooth flux. This boxcar smoothing was only used to remove outlying epochs, and was not used in our starspot analysis. These smoothing values were arrived at by eye to remove the most dramatic flares and outliers in the data. As this was not a comprehensive outlier removal scheme, some small amplitude flares and data systematics remained in the light curve. These small amplitude excursions occurred stochastically throughout the light curve, had no dependence on rotational phase, and therefore did not affect our spot modeling results. As discussed in \citet{lurie2015}, saturation can affect the \Kepler light curves for flare stars during the brightest flare events. However, the starspot modulations for GJ 1243 were low amplitude, and the quiescent flux level was not near the saturation limit. While the brightest flux excursions due to flares may be affected by saturation, our starspot analysis is not.
Our final, inter- and intra-quarter polynomial detrended, flare-cleaned light curve for GJ 1243 contained 47,478 epochs of data over the four years of \Kepler long cadence observations.

\subsection{Periodic Signal}
\label{sect:period}

The rapid rotation of GJ 1243 was first detected from periodic flux modulations due to starspots by \citet{irwin2011} using ground-based photometry from the MEarth project \citep{nutzman2008,irwin2009}. Following the initial Q0 release of \Kepler data, \citet{savanov2011} published the first analysis of the starspots on GJ 1243, using 44 days of continuous long cadence data. They reported a rotation period of 0.593 days for GJ 1243, and found that GJ 1243 exhibited two starspot features, separated in longitude by 203$^\circ$, and both stable in position over the 44 days of observation (equal to $\sim$74 rotation periods). The starspots covered 3.2\% of the visible stellar surface, with a modest amount of growth reported over Q0.

A study searching for rotation periods using the autocorrelation function for $\sim$2500 \Kepler M dwarf stars was carried out by \citet{mcquillan2013} using 10 months of \Kepler photometry. They reported a rotation period of 0.593 days for GJ 1243 as well. However, a larger scale analysis of over 40,000 active \Kepler stars by \citet{reinhold2013}, using the Lomb-Scargle periodogram method \citep{lomb1976,scargle1982}, did not report a rotation period for GJ 1243, as the star's rapid rotation was below their period cutoff.

These previous studies of GJ 1243 only reported the stellar rotation period to an accuracy of 0.001~day ($\sim$86 seconds). With such a short rotation period for GJ 1243, an error of 0.001 days would result in phase-folded data being out of phase by an entire rotation within one year. Thus, to measure any real phase evolution of the starspot features over four years we must determine the most accurate mean rotation period possible. We computed the normalized Lomb-Scargle periodogram using the entire 4-year detrended long cadence light curve, using no frequency oversampling or smoothing. The strongest peak in the resulting periodogram was very narrow, and had a period of 0.592596 days. We then computed the Lomb-Scargle periodogram over each of the 14 quarters of data individually. The mean period from all quarters we recovered was 0.592673 days, which was only $\sim$6.5 seconds longer than the period found from all quarters simultaneously. These 14 period estimates had a standard deviation of 0.00021 days, or about 18 seconds, which we adopt as the period uncertainty. Since the rotation period is very stable over the course of the \Kepler observations, we assume the period determined from the entire light curve, $P = 0.592596\pm0.00021$ days, for our analysis. 


\begin{figure}[]
\centering
\includegraphics[width=3.5in]{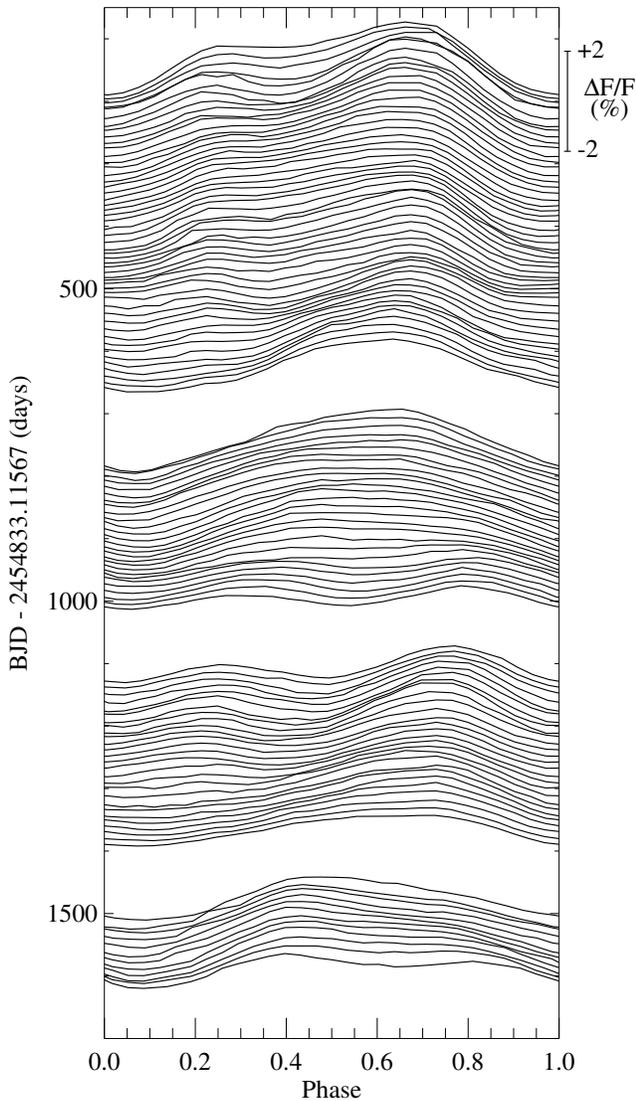}
\caption{Phase-folded, median smoothed light curves for GJ 1243 from 10-day windows of time, showing the slow evolution of the starspot modulations over time. The vertical position for each curve corresponds to the start time of the 10-day window on the left axis. Each time window is scaled to the same relative flux, shown on the right axis. The primary dip, centered at Phase=0 corresponds to the long-lived starspot.}
\label{phase}
\end{figure}

We then empirically defined the ephemeris of the flux minimum by phase-folding the entire \Kepler light curve at this rotation period. The phase of flux minimum was fit using a least squares regression with a Gaussian function, which determined an ephemeris of $t_0 = 2454833.11567807 \pm 0.00015$. 
In Figure \ref{phase} we show median-smoothed 10-day windows of the entire GJ 1243 light curve, phased using this rotation period and ephemeris. The primary dip in brightness stays fixed near Phase=0 over the 4 years of observation, which is due to the primary starspot. Slow evolution in both phase and amplitude of the secondary starspot feature is clearly seen. The secondary starspot is almost entirely absent at Time $\sim$ 700 days (using units of time as BJD - 2454833.11567 days), while the primary and secondary starspots appear to have nearly equal amplitudes at Time $\sim$ 1100.

\begin{figure*}[]
\centering
\includegraphics[width=7in]{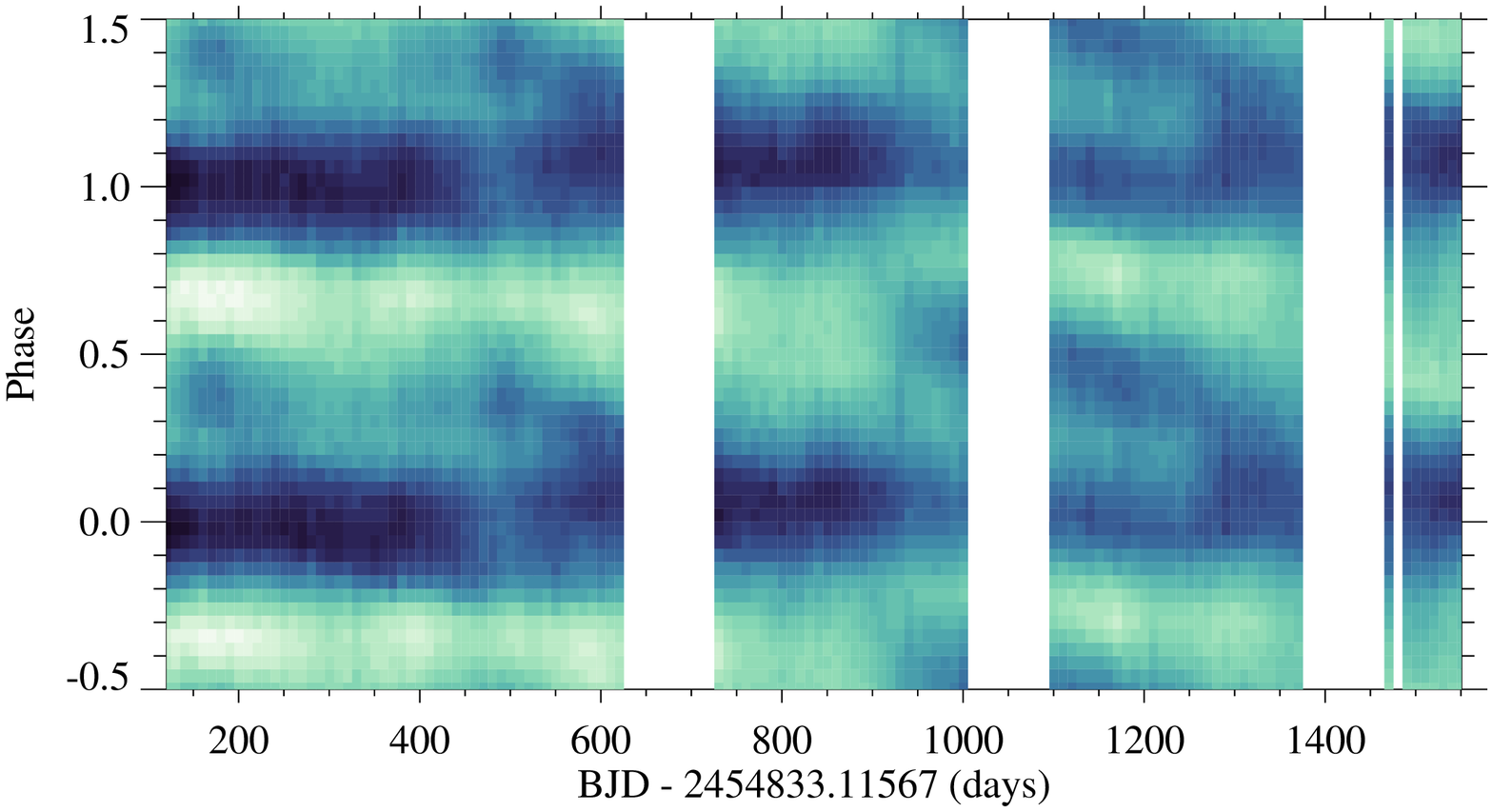}
\caption{Continuous phased light curve map for the entire \Kepler long cadence dataset. Pixel shade, from dark to light, indicates the median flux in each (time, phase) bin. Vertical white gaps correspond to times with no \Kepler data, as in Figure \ref{lc}. Pixels span 10 days in time and 0.04 in phase. The starspots are seen as dark regions in this diagram, which evolve in time from left to right.}
\label{lcmap}
\end{figure*}

Given the long starspot evolution timescale and short rotation period for GJ 1243, along with the nearly continuous \Kepler light curve for most of the 4-year timespan, we are able to study the change in starspot properties in much higher temporal detail than illustrated in Figure \ref{phase}. Using 10-day windows of time, we show the 4 year continuous phase evolution of flux from GJ 1243 in Figure \ref{lcmap}. For visual clarity the data is folded twice in phase. White vertical gaps correspond to quarters with no \Kepler data, as seen in Figure \ref{lc}. Each column of pixels in this phase versus time flux map contains data spanning 10 days. This binning resulted in $\sim$16 rotation periods per column, with an average of over 400 data points. Each row spans 0.04 in Phase, or equivalently a 14.4$^\circ$ slice in longitude. The median value for the flux within each (time, phase) pixel corresponds to the shading, with the darkest regions corresponding to a flux 1.5\% below the median value, and the lightest pixels 1.5\% above the median flux.

The dark band centered at Phase=0 in Figure \ref{lcmap}, which extends throughout the timespan of the data, is due to the primary starspot. This feature does not significantly change in phase over the course of our data. The flux amplitude for the primary spot is also nearly constant. There is an apparent change in the starspot flux amplitude around day $\sim$500 and day $\sim$900, due to the the presence of the secondary starspot combined with the systematic errors in the flux calibration.
The starspot features seen in Figure \ref{lcmap} are very large compared to spots seen on the Sun, appearing to span 50--90$^\circ$ in longitude. The detailed geometries of these features cannot be determined from this phase versus time flux map, and each observed ``starspot'' may in fact be a large spot group. Additionally, we cannot constrain the total starspot coverage, which may include many smaller spots and active regions across the entire stellar surface. Instead we are observing the total flux asymmetry due to these spots or spot groups.

The secondary starspot feature continuously changes in both phase and flux amplitude (equivalently pixel shade) in this diagram. 
This secondary feature seems to emerge and decay at least three times over the 4-year dataset, each time appearing nearly on the opposite hemisphere of the star and evolving towards the primary starspot.  Note, a decrease in phase corresponds to a starspot {\it advancing} in longitude in the direction of rotation over time. We interpret the slow, linear phase evolution of the secondary starspot to be the signature of differential rotation.

\section{Modeling the Light Curve}
\label{sect:mcmc}

To quantitatively trace the differential rotation on GJ 1243, we must determine the precise sizes and positions of the starspots over time. To accomplish this we performed a detailed fit to the \Kepler light curve using the starspot modeling software from L. Hebb (2015 in preparation). Here we give a brief overview of this light curve modeling program, as well as our specific use with the GJ 1243 system.

The starspot modeling code simulates the star as a sphere with uniform surface brightness and limb darkening onto which circular, gray starspots are fixed. Limb darkening is implemented by treating the star as a series of overlapping, concentric circles with brightness values defined by the 4-coefficient limb darkening model of \citet{Claret2011}. Starspots are modeled as non-moving circular regions with a fixed flux contrast relative to the photosphere, and may be placed anywhere on the stellar surface. At each time step in the input light curve, as the model star rotates, the code calculates the flux blocked by the spots rotating in and out of view, and thus generates a synthetic light curve.

The program can generate a synthetic light curve for a single star, with or without a transiting exoplanet,  and with the spin-axis of the star and orbital axis of the planet in any orientation (aligned or misaligned).    
To derive the properties (latitude, longitude and radius) for a number of spots that best reproduces the observed flux modulations, a $\chi^{2}$ comparison is made between the observed data and a synthetic light curve. 
The model light curve generating engine is wrapped with several types of Markov Chain Monte Carlo (MCMC) samplers, including an affine invariant MCMC based on \citet{emcee}, which explores the parameter space to find the lowest $\chi^2$, and thus the optimum spot properties.

The program requires that we choose the number of spots on the star a~priori, and that the spot distribution remains static. We only analyze a subset of the \Kepler data at any one time, using a ``window'' to model a timescale over which we do not expect the spots to evolve. By sliding this window over the full length of the light curve and running the code many times, we fit the entire light curve and determine the evolution of the spots. We emphasize that the MCMC runs are done independently, generating a unique best-fit spot solution within each window. This approach of multiple discrete models over time avoids parameterizing the starspot evolution with analytic functions as has been done previously \citep[e.g.][]{kipping2012}, which in turn allows us to track non-linear behavior in the size and position evolution of the spots. 
We refer the reader to L. Hebb (2015 in preparation) for a description of the full details and capabilities of this program, and briefly describe our specific use below.

We split the GJ 1243 light curve into windows with 5 day durations, or approximately 8.4 rotation periods at the \Kepler 30-minute cadence. The short rotation period, combined with the slow evolution of spot features seen in Figure \ref{lcmap}, resulted in many stellar rotations for each window, minimizing the effect of spurious light curve features such as flares or small data gaps. Each time window was required to contain at least 100 data points, or equivalently $\sim$3.5 rotation periods. Each subsequent time window was advanced by 2.5 days, providing two independent MCMC solutions for each datum.  A total of 447 such time windows were used spanning the 14 quarters of data.

We assumed a fixed flux contrast value of 0.7 for the starspots, which is consistent with contrast values seen for spots on active giants, as well as the average contrast of the solar umbra \citep[e.g.][]{berdyugina2005}. Note that while resulting spot sizes are directly dependent on the contrast value used in our model, the longitude and therefore the differential rotation is not affected.
The default value of 100 annuli was used to compute the limb darkening. Based on v~sin~i measurements from echelle spectroscopy of GJ 1243 (Wisniweski 2015 in preparation) and our measured rotation period, we used a fixed inclination of 32 degrees.

For each of the 447 windows of time, we modeled the GJ 1243 light curve using two starspots where each starspot is defined by fitting three parameters: its latitude, longitude and radius. This was the simplest model that was able to reproduce the observed flux modulations for all time windows to high accuracy. We note that some time windows were well fit using a single spot solution, particularly at Time $\sim$ 800 in Figures \ref{phase} and \ref{lcmap}, where the flux modulation was dominated by a single sine-curve like feature. Models with higher numbers of spots (three or more) were tested and could easily reproduce the observed flux modulations, but were not preferred when properly compared to the two-spot models with fewer free parameters.

Constraining the latitudes for large starspots is often difficult when deriving 2-dimensional starspot configurations from 1-dimensional light curves.  There exists a well known degeneracy between spot latitude and radius, resulting in families of solutions for spots at a given {\it longitude} but a range of latitudes and spot sizes that provide equally good fits to the observed light curve.  
Therefore, we chose to fix the latitudes for the two starspots to break this degeneracy in our model runs.   This does not affect our final conclusions because the differential rotation measurements depend only on the derived longitudes of the spots. 
To select latitudes at which to fix the two spots in our model, we ran our entire light curve modeling analysis for 1/10th of the time windows, and using five configurations of starspot latitudes. 
For each model configuration one spot was fixed at the stellar equator (0$^\circ$), and one at a higher latitude towards the inclined pole. The five higher latitudes spot positions tested were (72.8$^\circ$, 55.6$^\circ$, 38.4$^\circ$, 26.9$^\circ$, 9.8$^\circ$).  
Note, given the inclination of 32$^\circ$, spots above $\sim$58$^\circ$ would be partially or fully visible during the entire stellar rotation, and therefore would produce less flux modulation. The starspot longitudes and sizes were allowed to vary in configuration.

Each resulting set of MCMC solutions produced comparably good fits to the light curve, and had the same number of free parameters. The individual sine-like modulations seen in the light curve were not required to correspond to the higher or lower latitude spot in any given model configuration. As a result, some models would exhibit a ``flip'' between spot latitudes for a given feature at nearly the same longitude between subsequent time windows. This flipping was observed for the two configurations with higher latitude spots (72.8$^\circ$ and 55.6$^\circ$). For our final analysis we chose the solution set with the highest latitude configuration that did not exhibit this flipping in spot latitudes between subsequent time windows. The two starspots in our analysis were therefore fixed at latitudes of 38.4$^\circ$ and 0$^\circ$. We note our resulting longitudinal shear results were insensitive to the latitudes chosen.

The affine invariant sampler based on \citet{emcee} was employed for each time window, with random starting values for the spot radius and longitude, but fixed latitudes as described above. 
For each window of data, the MCMC was run for 300 steps using 100 walkers, and the ``a scale'' parameter was set to 2.0. To carry out these independent MCMC realizations efficiently in parallel, we used {\tt CONDOR} \citep{condor-hunter,condor-practice} to distribute the 447 MCMC explorations across 180 Linux workstation computer cores.  Each window's MCMC chain was converged after 300 steps, and the starspot configuration that produced the best-fit
(lowest $\chi^2$) solution for each time window was adopted. 
We note that the phase-folded data within each window had a scatter about 10 times greater than the typical photometric uncertainty given in the \Kepler data. This was due to errors in the underlying light curve and limitations of our detrending algorithm, as well as small amplitude evolution of the starspot features within each window. The starspot flux modulation signal was more than 20 times greater than this scatter. 
Average reduced $\chi^2$ values were $\sim$2 per time window, assuming a 10 times increase in the photometric uncertainty.

\begin{figure*}[!h]
\centering
\includegraphics[width=6.5in]{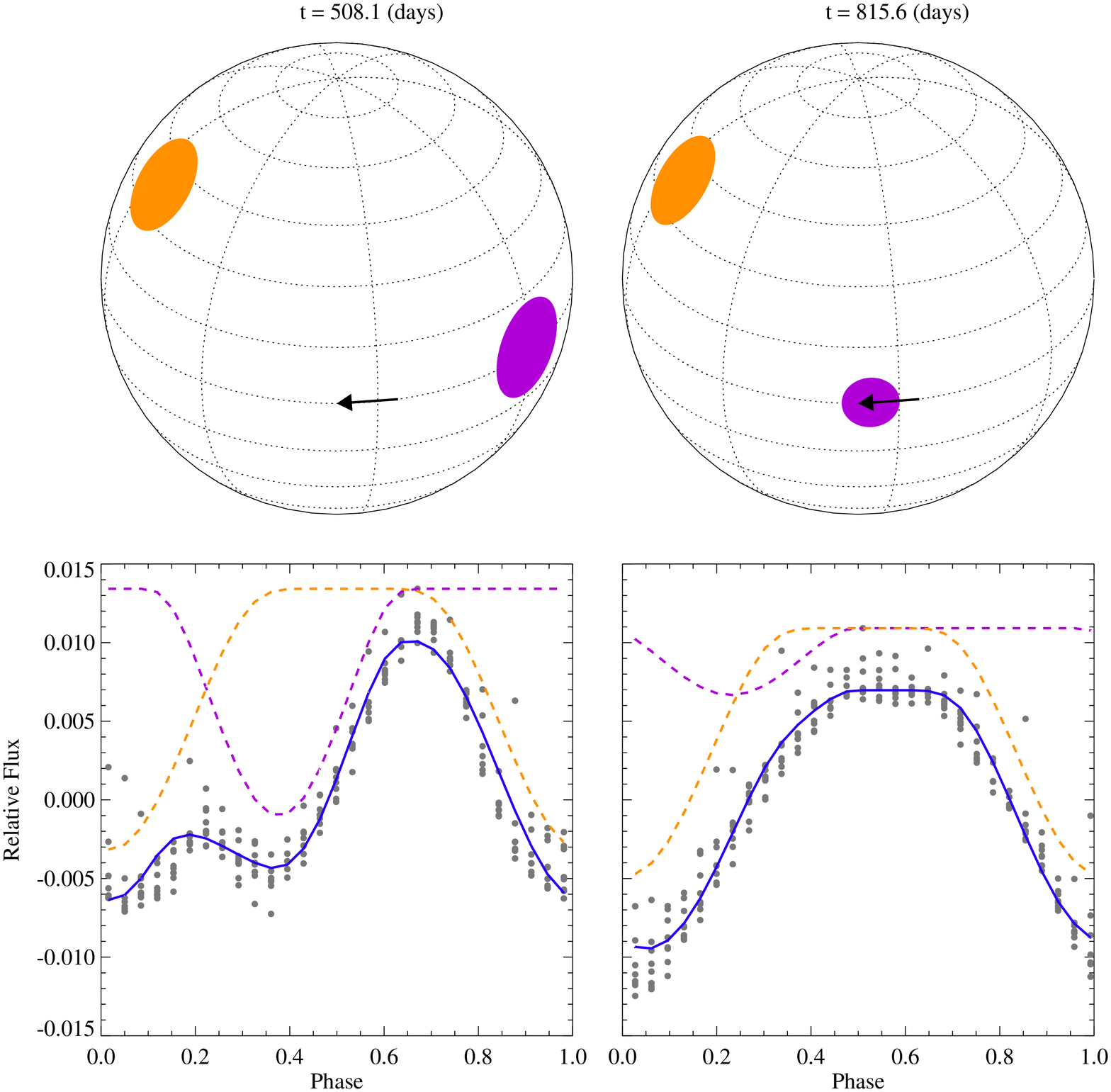}
\caption{Top: orthographic projections of the model star, with an inclination of 32$^\circ$, and the best-fit positions for two circular spots for the 5 day time window starting at BJD - 2454833.11567 = 508.1 days (left) and 815.6 (right). The direction of stellar rotation is indicated by the black arrow. Bottom: phase-folded light curve for the data in the same 5 day time windows, with the best-fit two-spot models overlaid (blue solid line), and the contributions from both the higher latitude (orange dashed line) and equatorial (purple dashed line) starspots offset for clarity.}
\label{map}
\end{figure*}

In Figure \ref{map} we show phase-folded light curves for two representative time windows of data with their best fit solutions overlaid, along with orthographic map projections of the model stellar surface showing the best-fitting spot configurations. This map projection demonstrates the inclination of the star as well as the relatively large size of the starspots. The higher latitude spot appears at nearly the same phase (longitude) and size, while the lower latitude spot shrinks in radius and advances in longitude (lower phase) in the direction of the stellar rotation.

\begin{figure*}[!t]
\centering
\includegraphics[width=7in]{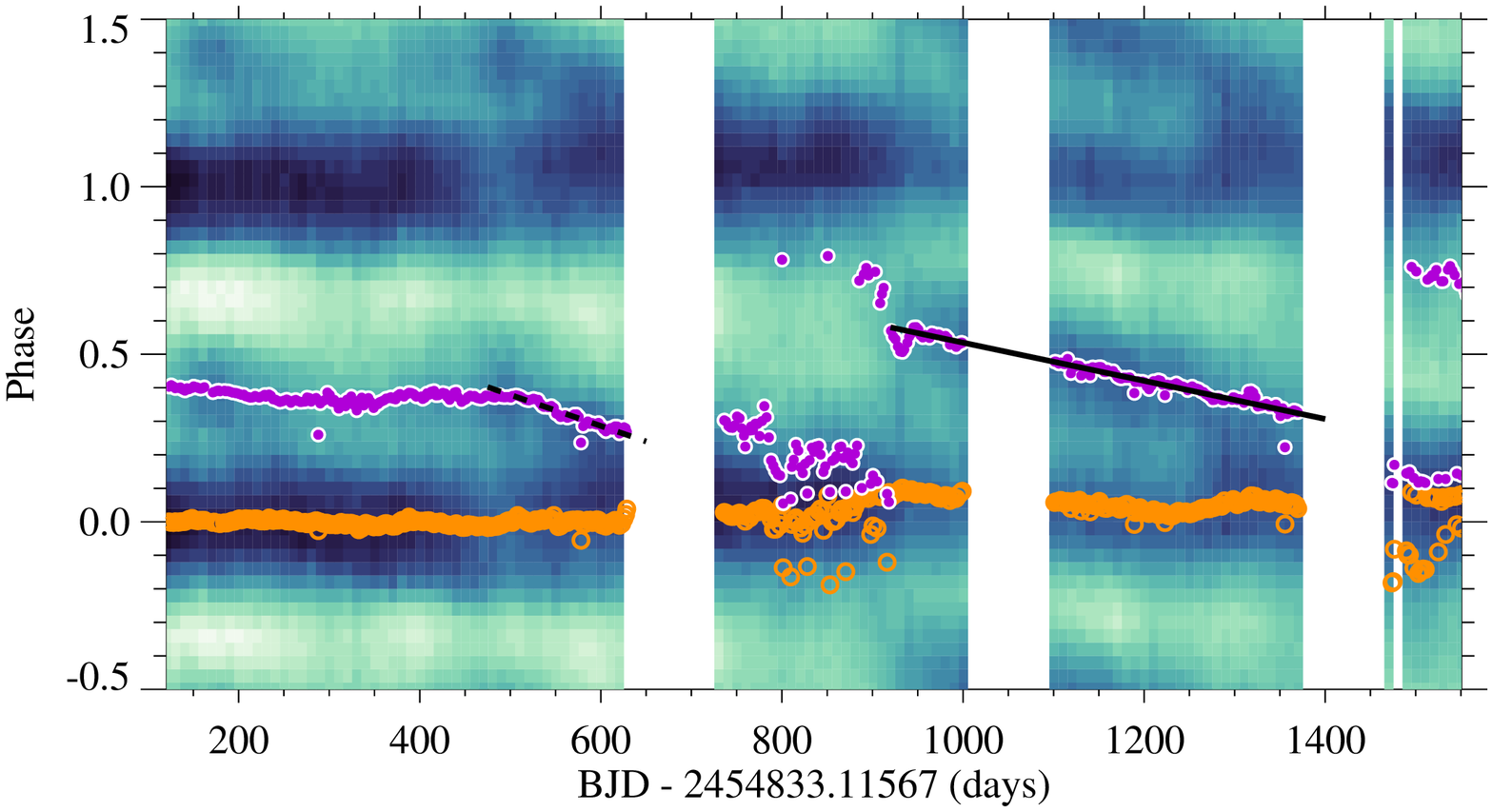}\\
\includegraphics[width=7in]{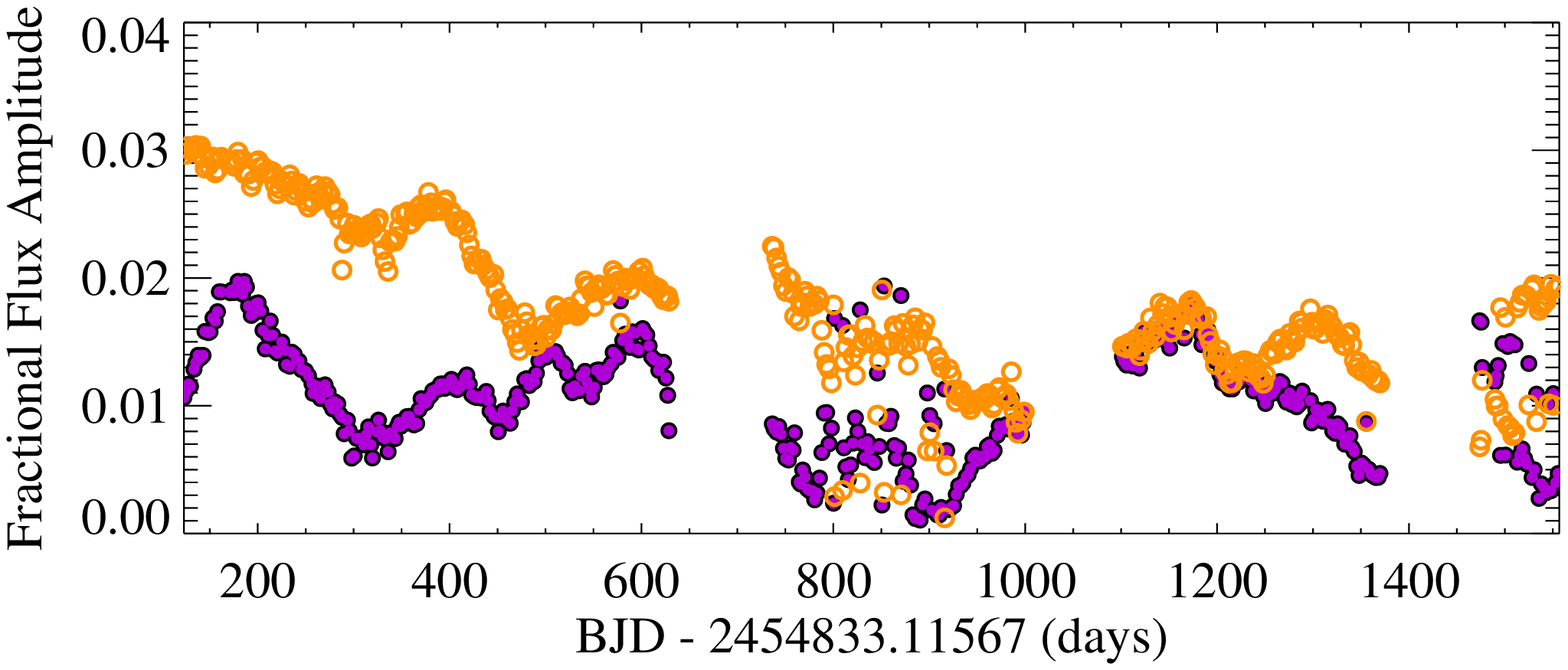}
\caption{
Top: Continuous phased light curve map, as in Figure \ref{lcmap}, with the best-fit solutions from our two spot model overlaid. The higher latitude spot shown in Figure \ref{map} (orange open circles) remains nearly constant in phase, while the secondary lower latitude spot (purple filled circles) evolves significantly. Linear fits to the phase evolution for the secondary spot are overlaid (black solid and dashed lines), which we interpret as differential rotation. 
Bottom: Fractional flux amplitude of each starspot as a function of time for the best-fit solutions from our two spot model. Colors are the same as above.
}
\label{diffrot}
\end{figure*}

\section{Quantifying the Starspot Evolution}
\label{sect:trace}

Using the best-fit parameters from each stationary, independent MCMC model, as in Figure \ref{map}, we were able to trace the sizes and longitudes of two starspots over the entire span of our \Kepler data. In Figure \ref{diffrot} we show the rotational phase (equivalently the longitude facing the observer) for both starspots as a function of time. 
The higher latitude starspot indicated in Figure \ref{map} (orange) is very stable in Figure \ref{diffrot} in phase (longitude) with a standard deviation in longitude of only 4.5\% (16 degrees) over the four years of data, and traces the dark band seen in Figure \ref{lcmap} centered at Phase = 0. The amplitude of this higher latitude spot on the light curve changes slowly over the data, with a standard deviation of 34\% in fractional flux in Figure \ref{diffrot}. We refer to this feature as the ``primary starspot''.

The ``secondary starspot'' (purple), however, evolves significantly in phase across the stellar surface over time in Figure \ref{diffrot}. This feature corresponds to the lower latitude, equatorial starspot in Figure \ref{map}, and traces the transient secondary features seen in Figure \ref{lcmap}. Between Time $\sim$750 and 900, the two best-fit starspot locations were very close in phase, and the variance between solutions in subsequent time steps increased for both the primary and secondary spots. These correspond to time windows where a one-starspot model would be preferred. 

We manually identified two regions in Figure \ref{diffrot} that displayed nearly constant linear evolution in the secondary starspot longitude: Time = 510--630, and Time = 945--1400. We interpret these to be the signatures of differential rotation, with secular spot motions in time. Within these time windows we used a non-linear least squares first order polynomial fit to measure the linear slopes. Lines of best fit for these two regions are shown in Figure \ref{diffrot} as dashed and solid black lines, and had slopes of -0.000927 and -0.000569, respectively. 
The occurrence of these secondary starspots at multiple times within our data may in fact be due to a single lower latitude feature lapping the primary starspot, but we note the slopes and separations in these features in Figure \ref{diffrot} are not consistent with a single spot at a fixed rate of differential rotation.

The measured slopes were in units of phase day$^{-1}$, and corresponded to a rotation shear of $\Delta \Omega = 2\pi/t_{lap}$ = 0.0058 and 0.0036 rad day$^{-1}$, using the definition from \citet{kuker_rudiger2008}. Note however this does not include any consideration of the starspot latitudes. These slopes can also be converted to secondary rotation periods using the equation
\begin{equation}
P_i = \dfrac{P_0}{1 - m_i P_0} \, ,
\end{equation}
where $m_i$ is the slope of each feature, $P_0$ is the rotation period used to phase fold the data to make the figure, and $P_i$ is the resulting rotation period. Note by this definition a negative slope yields a shorter rotation period.

Differential rotation is generally parameterized \citep[e.g.][]{henry1995} as:
\begin{equation}
P_\phi = P_{eq} / (1 - k \sin^2\phi) \,,
\end{equation}
where $P_\phi$ is the rotation period at a given latitude ($\phi$), $P_{eq}$ is the rotation period at the equator, and $k \equiv \Delta\Omega / \Omega_{eq}$ governs the rate of differential rotation as a function of latitude. Our model results indicate that the period used to phase fold the data in Figure \ref{diffrot} corresponds to the higher latitude (38.4$^\circ$) starspot, and assumes the secondary starspot features are on the stellar equator. Using an average slope from Figure \ref{diffrot} of $m = -0.000748$, and the phase-folding period from \S2.2, we estimated an equatorial rotation period of $P_{eq} = 0.5923336$ days via Equation 1. We then solved for the unitless differential rotation parameter using Equation 2, finding $k = 0.00114$, which corresponds to $\Delta\Omega = 0.012 \pm 0.002$ rad day$^{-1}$. The uncertainty we quote here is propagated from the errors in the linear least squares fits in Figure \ref{diffrot}.

Assuming the primary spot for GJ 1243 is indeed at a higher latitude than the faster rotating secondary spots, this behavior is consistent with Solar-like surface differential rotation where the equator rotates faster than the poles. If the primary and secondary starspots are well separated in latitude as our model indicates, such a low value of shear indicates very weak differential rotation, with the star rotating nearly as a solid body. For comparison, the Sun's surface differential rotation is much stronger, with $\Delta \Omega$  = 0.055 rad day$^{-1}$ \citep{berdyugina2005}.

\section{Fitting with Gaussians}
\label{sect:gauss}

\begin{figure*}[!t]
\centering
\includegraphics[width=7in]{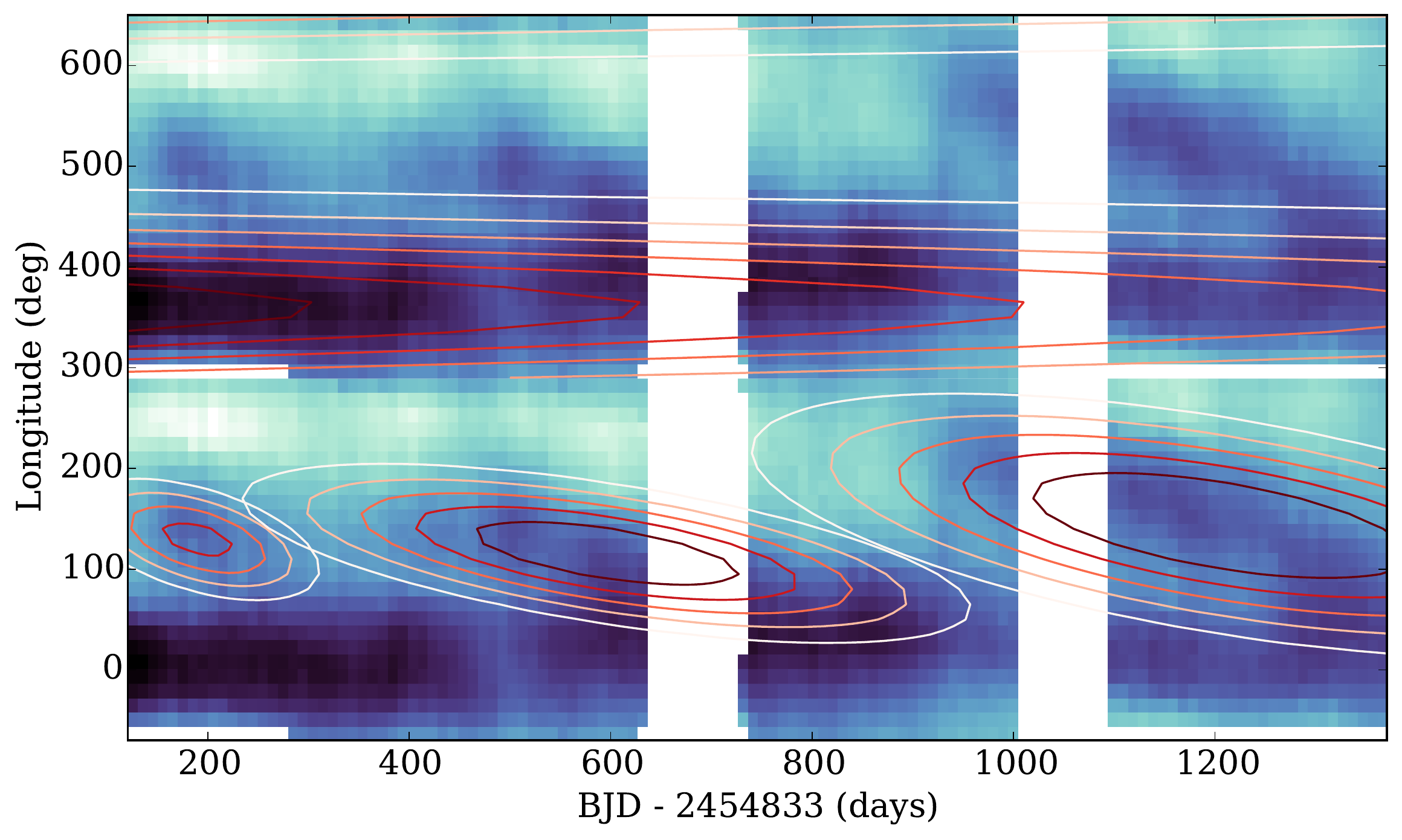}
\caption{Bivariate Gaussian models of starspot evolution in phase and time (open contours), overlaid on the time--phase flux map from Figure \ref{lcmap}. A total of four Gaussians defined by Eqn \ref{gfunc} were fit, representing one primary and three secondary starspot features. For visual clarity we have offset the Gaussian that corresponds to the primary starspot by 1 phase.
}
\label{pymcmc}
\end{figure*}

In the previous two sections we have focused on measuring starspot evolution using a series of sophisticated stationary models, and finding the differential rotation rate by comparing the position of spots in subsequent model realizations. In this section we explore an alternative method of explicitly determining the starspot time evolution, and thus the differential rotation rate, using Gaussian functions. 

Rather than modeling the entire light curve directly to infer the starspot sizes and positions, as in Figure \ref{map}, we analyzed the three-dimensional ``surface'' shown in Figures \ref{lcmap} and \ref{diffrot}, which traces the flux as a function of both time and rotational phase. The data were binned in both time and phase (longitude), using bin sizes of 10 days and 14.4 degrees, respectively. To model this flux map we used 2-dimensional bivariate Gaussian distributions of the form:

\begin{eqnarray}
t' = (t-t_0) \cos \theta - (l-l_0) \sin\theta \notag \\
l' = (t-t_0) \sin \theta - (l-l_0) \cos\theta \notag \\ 
F(t,l) = A  \exp \left( - \frac{(t'/\tau)^2 + (l'/b)^2}{2} \right)
\label{gfunc}
\end{eqnarray}

\noindent
where $F(t,l)$ is the flux as a function of time $t$ and longitude $l$, $A$ is the flux amplitude of the starspot, $\tau$ is the lifetime of the starspot, $b$ is the scale width of the large starspot in longitude, $t_0$ is the center time of the starspot, $l_0$ is the center longitude of the starspot, and $\theta$ is the slope of the spot evolution in units of degrees day$^{-1}$. Here the longitude is a circular coordinate, with range between 0 and 360 degrees, and defined to continuously wrap from 360 back to 0 degrees. This definition enables long-lived starspots with large rates of differentially rotation to ``lap'' the stellar surface multiple times. Each starspot's evolution is defined by evaluating Equation \ref{gfunc} over the entire time span of our data, and the full range in longitude. The entire flux map in Figure \ref{map} is reproduced by summing  the 2-D Gaussian functions.

For our Gaussian analysis we used the data spanning from the beginning of the available \Kepler data (Q0) through Quarter 14 (Time $\sim$ 1370). This time range was chosen to focus our analysis on the secondary spot evolution and differential rotation signal measured above.  We discarded data from Q16 and Q17 data which showed no sign of the secondary starspot.

We solved for the positions and evolution of four starspots over the entire duration of the data, using the Python MCMC sampler {\tt emcee} from \citet{emcee} to explore parameter space for all four simultaneously. In this four Gaussian model, we consider the largest spot (also with the smallest slope in $\theta$) as the primary, and the three secondary spots as independent spot features, or repeat occurrences of the secondary spot discussed before. 
A third occurrence of the secondary spot feature was needed to account for the small feature seen around Phase$\sim$0.4 at Time$\sim$100 days in Figure \ref{lcmap}, which was not chosen in our conservative by-eye selection above.
We used rough values for the parameters to seed {\tt emcee} with. The initial seeds for the primary spot were $\theta=0$, $t_0=800$, $\tau = 1300$ days, $l_0=75$ degrees, and $b=15$. The three secondary spots were all seeded with $\theta$ = -0.2 degree day$^{-1}$, $\tau = 100$ days, $l_0=200$ degrees. The secondary starspot center times seeds were set to $t_0$ = 200, 530, and 1200.

We then ran {\tt emcee} with 50 walkers for 2000 steps. The best fit model from this parameter space search for the three secondary starspots is shown in Figure \ref{pymcmc}. The best fit slope for the primary spot was $\theta$ = -0.0011 deg day$^{-1}$, and for the secondary spots (in time order) was $\theta$ = -0.20, -0.14, and -0.15 deg day$^{-1}$. These secondary spot shear rates corresponded to $2\pi / t_{lap} = \Delta \Omega$ = 0.0036, 0.0025, and 0.0027 rad day$^{-1}$, 
or an average of 0.0029 rad day$^{-1}$, somewhat lower amplitude than measured from the linear features in Figure \ref{diffrot}. Note again this method does not constrain the latitudes of the starspots, and so the measurement of shear is only a lower limit on the true differential rotation rate.

This approach assumes a priori that the starspot evolution in both time and longitude can be represented by a Gaussian function, meaning the spots may only evolve linearly in longitude over time. We note the resulting estimate for the differential rotation shear rates for the secondary starspots are very similar to the values determined when fitting many time-stationary MCMC instances. The Gaussian modeling approach used flux data that was binned in time and phase, greatly reducing the number of data points to be fit. This entire MCMC analysis took only a few minutes to compute using a standard Linux workstation. We thus propose this to be an efficient means of estimating the starspot lifetimes and differential rotation rates in certain cases, which we discuss further in Section \ref{sect:discussion} below.

\section{Discussion}
\label{sect:discussion}

We have produced two significant results in this work. 
The first is the identification of a long-lived starspot, which we attribute to a higher latitude starspot (possibly due to a spot cap or group) on the rapidly rotating M dwarf, GJ 1243. The second is a robust measurement of weak differential rotation for this star due to a spot closer to the stellar equator. 
In this section we provide additional context and discussion of these results, and their implications for the magnetic field's surface topology.

\subsection{A Long Lived Starspot}

To further illustrate the remarkable stability of the higher latitude spot on GJ 1243, we retrieved ground-based light curves from SuperWASP which predated the \Kepler mission by $\sim$2 years \citep{butters2010}. This SuperWASP public archive photometry was phased using the period and ephemeris we determined from our \Kepler light curve. The phase of flux minimum matches between these two datasets to within 1\%, indicating this large starspot has been stable in longitude for more than 6 years. The amplitudes of the flux modulations between \Kepler and SuperWASP are only slightly different, with the median \Kepler variation of 2.19\% (averaged over all 4 years of data), and for SuperWASP of 2.86\%. The SuperWASP data was taken in the V-band, which is more narrow than the very wide \Kepler filter. The V-band also is centered at a shorter wavelength than the \Kepler filter, which is weighted more towards the R-band.
As a result, we would expect to find larger flux contrast between cool spots and the stellar photosphere in the V-band than compared to the \Kepler filter. However, since these observations were not concurrent we cannot rule out small differences in the starspot's physical size over time.

\subsection{Differential Rotation in Cool Stars}
\begin{figure}[!t]
\centering
\includegraphics[width=3.5in]{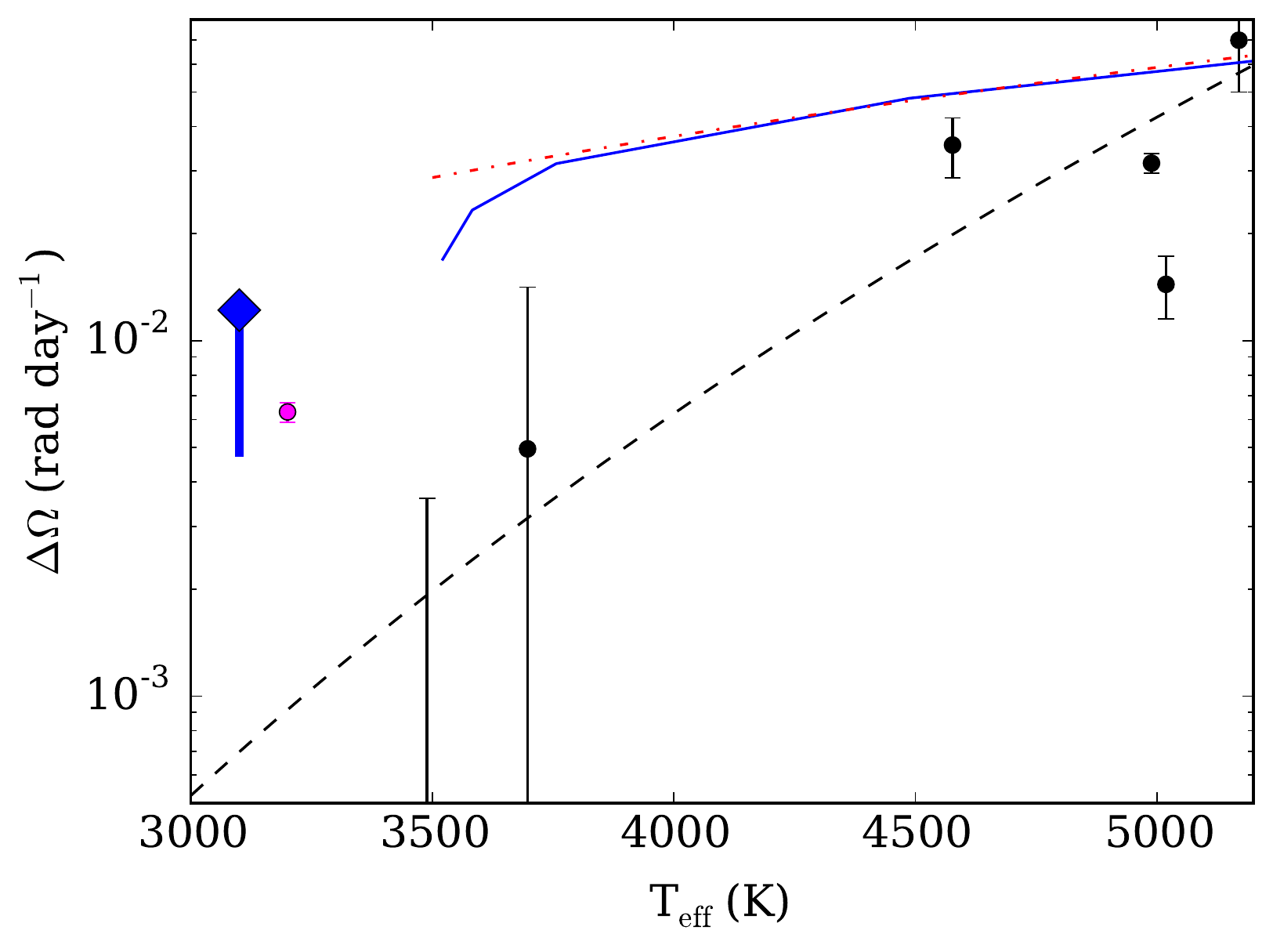}
\caption{Average starspot shear from the two linear fits to the MCMC light curve models in Figure \ref{diffrot}, assuming the primary spot was at a latitude of 38.4$^\circ$ and secondary spot at the equator (blue diamond). The blue bar extends to the minimum possible amplitude of differential rotation for GJ 1243, assuming the primary and secondary spots are at the pole and equator, respectively. For comparison, the \citet{cameron2007} observed fit for cool stars (black dashed line), theoretical prediction from \citet{kuker_rudiger2011} (red dot-dash and blue solid lines), individual stars from \citet{barnes2005a} (black circles), and the estimated shear rate for V374 Peg determined using doppler imaging from \citet{morin2008a} (purple filled circle) are shown.}
\label{dr}
\end{figure}

The average starspot shear for GJ 1243 observed in this paper of 0.0047 rad day$^{-1}$ corresponds to a differential rotation rate of $\Delta\Omega = 0.012 \pm 0.002$ rad day$^{-1}$ (assuming the spot configurations used in our models),
and is one of only a few such measurements yet obtained for low mass, rapidly rotating, fully convective stars. In Figure \ref{dr}, we place this measurement in the context of other existing observations of stellar differential rotation \citep{barnes2005a,morin2008a} along with the empirical extrapolation to cool stars from \citet{reiners2006,cameron2007}, and models from \citet{kuker_rudiger2011}. One of the few other objects with a robust differential rotation measurement in this regime is the cool, rapidly rotating star, V374 Peg \citep{morin2008a}. These authors employ Doppler Imaging, a completely different technique to our own, to derive a value for the surface differential rotation of V374 Peg ($\Delta\Omega = 0.0063 \pm 0.0004$ rad day$^{-1}$) that is similar to that of GJ 1243 we have measured. 
This Doppler Imaging method assumes a Solar-like differential rotation profile as in Equation 2, and simultaneously fits for the starspot positions, sizes, and shear rates.
\citet{lurie2015} have also estimated the 
starspot shear 
for both rapidly rotating components of the M5+M5 binary GJ 1245AB, using \Kepler data and the 2D Gaussian modeling approach detailed in our Section \ref{sect:gauss}, finding shear rates that are comparable to GJ 1243 and V374 Peg.

The various methods for constraining differential rotation each have unique limitations in their sensitivity and degeneracies. Photometric phase-tracking methods such as ours, frequency splitting approaches \citep{reinhold2013}, as well as spectroscopic line broadening techniques \citep[i.e. the Fourier Transform Method; ][]{reiners2003} have been considered as only providing lower limits on the true differential rotation rate, due to a lack of constraint on the starspot latitudes \citep{barnes2005a}. Statistical corrections for this latitudinal uncertainty have been developed \citep{hall1994b}, and our models of GJ 1243 provide a weak constraint on the spot latitudes. 
Together, these varied observations indicate that rapidly rotating, mid M-dwarf stars exhibit differential rotation that is significantly weaker than on the Sun (0.055 rad day$^{-1}$) by up to an order of magnitude.

These observational results of low $\Delta\Omega$ values support recent theoretical work in this area.  
Mean field theory models predict that for stars at a fixed temperature surface differential rotation rates decrease with faster rotation (shorter periods), down to periods of a few days \citep{kuker_rudiger2005}. In addition, the total amplitude of the differential rotation decreases with decreasing temperature, as models indicate lower mass main sequence stars (T$_{\rm eff} < 6000$ K) should exhibit lower amounts of surface differential rotation than hotter stars at a fixed rotation period \citep{kuker_rudiger2011}.

Dipolar dynamos with strong magnetic field and quenched differential rotation have been reported in global dynamo models of rapidly rotating low-mass stars \citep[e.g.][]{gastine2013}. Furthermore, recent global dynamo modeling efforts have also been successful in producing polar starspots self consistently \citep{yadav2015}
Furthermore, in hydrodynamic simulations of rotating, non-magnetic, solar-type stars, \citet{browning2008} found that  convection in the outer zone of the star redistributes angular momentum quickly giving rise to solar-like differential rotation. However, when strong magnetic fields are introduced, the field lines act to reduce the differential rotation by linking together individual regions of the stellar interior.  As the strength of the magnetic field is increased, as is typically seen for more rapidly rotating stars, the differential rotation is suppressed to almost negligible values. 

The observations from \citet{lurie2015} also support this connection between rotation, magnetic field strength, and suppressed differential rotation. For the M5+M5 binary system GJ 1245AB, the more rapidly rotating component (GJ 1245A, P=0.26 days) has a slightly higher total chromospheric H$\alpha$ emission flux, and significantly less phase evolution of its starspot modulations compared with GJ 1245B, (P=0.71 days).

\subsection{Magnetic Field Topology}

Dynamo models and observations of stars like GJ 1243 are in agreement that for rapidly rotating, low-mass stars, the magnetic field strength is increased and surface differential rotation is suppressed.
The detailed surface topology of the magnetic field is less certain.
The convective dynamo models from \citet{browning2008}, for example, indicate that rapid rotation produces strong axisymmetric magnetic fields. Earlier models of convective envelopes  \citep{dobler2006} also show net axisymmetric magnetic fields. Other models of fully convective stars, however, have produced fully non-axisymmetric fields at the surface \citep{kuker_rudiger2005,chabrier2006}.

The smooth rotational flux modulation observed is due to a local feature on the stellar surface. The magnetic activity from this star as traced by flares in \Kepler data has also been well studied \citep{hawley2014,davenport2014b}. No correlation between rotational phase (or equivalently longitude) and the occurrence rate or energy emitted from  flares has been found. The star may be uniformly covered by many smaller active regions and spots that would not create observed modulations in the light curve. The flares would be a result of these small scale multipolar magnetic field structures. Similarly, the stellar pole may be entirely covered by a polar spot ``cap'', resulting from strong poloidal magnetic flux geometry due to the rapid stellar rotation. This poloidal component of the magnetic field could also be slightly misaligned from the stellar rotation axis, as seen in other convective stellar and planetary dynamos \citep[e.g.][]{christensen2009,hull2013}, resulting in the observed light curve asymmetries.

As we noted in Section 2, the Rossby number for GJ 1243 is $R_o\approx0.008$. 
According to the ZDI observations of M dwarfs aggregated in \citet{gastine2013}, for stars with $R_o < 0.1$ both dipole and multipole fields are possible. In this low Rossby number regime, they also find surface differential rotation should be stronger when the magnetic field is multipolar. However, we find for GJ 1243 a very low rate of differential rotation, and large, long-lived starspot modulations. As a result, in this context we predict a highly organized and stable dipolar magnetic field geometry. This is in agreement with the ZDI observations of the similar star V374 Peg, which has a large scale dipolar field and long-lived starspots \citep{morin2008a}.

\section{Summary}
\label{sect:summary}

In this paper, we have presented a classic approach of phase tracking starspots in light curves, made new by the exquisite photometric monitoring from \Kepler. By tracing the phase evolution for two starspot regions on GJ 1243 we have found the smallest amplitude of differential rotation rate ever robustly measured for a cool star. 
This phase-tracking technique is similar to \citet{henry1995}, and is sensitive to a comparable amplitude differential rotation signal to \citet{morin2008a}. The large starspots, or starspot groups, on GJ 1243 are very long-lived, with the primary high-latitude spot found to be constantly aligned in phase for over 6 years. The secondary starspot features evolve on timescales of hundreds of days in both phase and amplitude. 

There remain many challenges in modeling the starspots using broadband light curves alone. For example, we have almost no constraint on the actual latitudes of the spots. Modeling starspots on stars with transiting exoplanets \citep[e.g.][]{sanchis-ojeda2013} may help break many of these degeneracies. As most transiting systems in the \Kepler data are around G dwarf stars, more M dwarf systems like \Kepler 186 \citep{kepler186f} are needed to better understand the detailed starspot characteristics of stars across the main sequence.

Both the light curve modeling MCMC and 2D Gaussian phase-tracking techniques used to measure surface differential rotation in this work are best suited for tracking long-lived spots on rapidly rotating stars, as in the GJ 1243 system where we are able to average over hundreds of stellar rotation periods during a starspot's lifetime. We believe the light curve modeling approach provides the most robust estimates for starspot properties, but note the 2D Gaussian approach is orders of magnitude faster to execute.
This methodology could be applied to hundreds of rapidly rotating active stars in the \Kepler dataset for which rotation periods are already known \citep[e.g. see][]{reinhold2013,mcquillan2014}.
There are $\sim$20 other stars in the \Kepler data with M dwarf colors and estimated rotation periods shorter than 1 day \citep{mcquillan2013}. A cursory look at these light curves reveals many with dramatic flare activity and sinusoidal starspot modulations, as found on GJ 1243 and GJ 1245AB. The phase versus time diagrams (as in our Figure \ref{diffrot} for GJ 1243) for these other rapidly rotating stars show a diverse set of morphologies, ranging from even more stable spots than on GJ 1243, to stars with faster shear rates and shorter spot lifetimes.

Finally, we have introduced an efficient technique for empirically tracking starspot evolution in the phase versus time flux map, by fitting bivariate Gaussians to model the spot motion and evolution. Studying the phase-evolution of the starspots with \Kepler for single field stars appears to be feasible for stars with fast rotation periods and long spot lifetimes. 
The Gaussian-fitting method presented here has already been applied to the \Kepler data for the active M5+M5 binary system, GJ 1245 AB \citep{lurie2015}. 
We have pointed out many other rapidly rotating low-mass stars in the \Kepler archive that may be studied with this technique, and hope this work will be the beginning of a larger observational understanding of surface differential rotation in cool stars.

\acknowledgments
The authors wish to thank the anonymous referee, whose helpful comments greatly improved the quality of this manuscript. 
This work was supported by \Kepler Cycle 2 GO grant NNX11AB71G and Cycle 3 GO grant NNX12AC79G and NSF grant  AST13-11678. We are pleased to thank John Gizis for advance use of his GO program's long-cadence data on GJ 1243. We thank John P. Wisniewski and Adam F. Kowalski for continuing discussions on stellar activity, and Rakesh Yadav for very useful comments on an early draft of this manuscript. Davenport thanks Morgan Fouesneau for help with Python code.

Kepler was competitively selected as the tenth Discovery mission. Funding for this mission is provided by NASAÕs Science Mission Directorate.



\begin{thebibliography}{64}
\expandafter\ifx\csname natexlab\endcsname\relax\def\natexlab#1{#1}\fi

\bibitem[{{Aigrain} \& et~al.(2015)}]{aigrain2015}
{Aigrain}, S., \& et~al. 2015, \mnras, submitted

\bibitem[{{Barnes} {et~al.}(2005){Barnes}, {Collier Cameron}, {Donati},
  {James}, {Marsden}, \& {Petit}}]{barnes2005a}
{Barnes}, J.~R., {Collier Cameron}, A., {Donati}, J.-F., {James}, D.~J.,
  {Marsden}, S.~C., \& {Petit}, P. 2005, \mnras, 357, L1

\bibitem[{{Barnes} {et~al.}(2004){Barnes}, {James}, \& {Collier
  Cameron}}]{barnes2004}
{Barnes}, J.~R., {James}, D.~J., \& {Collier Cameron}, A. 2004, \mnras, 352,
  589

\bibitem[{{Berdyugina}(2005)}]{berdyugina2005}
{Berdyugina}, S.~V. 2005, Living Reviews in Solar Physics, 2, 8

\bibitem[{{Bertello} {et~al.}(2012){Bertello}, {Pevtsov}, \&
  {Pietarila}}]{bertello2012}
{Bertello}, L., {Pevtsov}, A.~A., \& {Pietarila}, A. 2012, \apj, 761, 11

\bibitem[{{Bonomo} \& {Lanza}(2012)}]{bonomo2012}
{Bonomo}, A.~S., \& {Lanza}, A.~F. 2012, \aap, 547, A37

\bibitem[{{Borucki} {et~al.}(2010){Borucki}, {Koch}, {Basri}, {Batalha},
  {Brown}, {Caldwell}, {Caldwell}, {Christensen-Dalsgaard}, {Cochran},
  {DeVore}, {Dunham}, {Dupree}, {Gautier}, {Geary}, {Gilliland}, {Gould},
  {Howell}, {Jenkins}, {Kondo}, {Latham}, {Marcy}, {Meibom}, {Kjeldsen},
  {Lissauer}, {Monet}, {Morrison}, {Sasselov}, {Tarter}, {Boss}, {Brownlee},
  {Owen}, {Buzasi}, {Charbonneau}, {Doyle}, {Fortney}, {Ford}, {Holman},
  {Seager}, {Steffen}, {Welsh}, {Rowe}, {Anderson}, {Buchhave}, {Ciardi},
  {Walkowicz}, {Sherry}, {Horch}, {Isaacson}, {Everett}, {Fischer}, {Torres},
  {Johnson}, {Endl}, {MacQueen}, {Bryson}, {Dotson}, {Haas}, {Kolodziejczak},
  {Van Cleve}, {Chandrasekaran}, {Twicken}, {Quintana}, {Clarke}, {Allen},
  {Li}, {Wu}, {Tenenbaum}, {Verner}, {Bruhweiler}, {Barnes}, \&
  {Prsa}}]{borucki2010}
{Borucki}, W.~J., {et~al.} 2010, Science, 327, 977

\bibitem[{{Browning}(2008)}]{browning2008}
{Browning}, M.~K. 2008, \apj, 676, 1262

\bibitem[{{Butters} {et~al.}(2010){Butters}, {West}, {Anderson}, {Collier
  Cameron}, {Clarkson}, {Enoch}, {Haswell}, {Hellier}, {Horne}, {Joshi},
  {Kane}, {Lister}, {Maxted}, {Parley}, {Pollacco}, {Smalley}, {Street},
  {Todd}, {Wheatley}, \& {Wilson}}]{butters2010}
{Butters}, O.~W., {et~al.} 2010, \aap, 520, L10

\bibitem[{{Chabrier} \& {K{\"u}ker}(2006)}]{chabrier2006}
{Chabrier}, G., \& {K{\"u}ker}, M. 2006, \aap, 446, 1027

\bibitem[{{Christensen} {et~al.}(2009){Christensen}, {Holzwarth}, \&
  {Reiners}}]{christensen2009}
{Christensen}, U.~R., {Holzwarth}, V., \& {Reiners}, A. 2009, \nat, 457, 167

\bibitem[{{Claret} \& {Bloemen}(2011)}]{Claret2011}
{Claret}, A., \& {Bloemen}, S. 2011, \aap, 529, A75

\bibitem[{{Collier Cameron}(2007)}]{cameron2007}
{Collier Cameron}, A. 2007, Astronomische Nachrichten, 328, 1030

\bibitem[{{Davenport} {et~al.}(2014){Davenport}, {Hawley}, {Hebb},
  {Wisniewski}, {Kowalski}, {Johnson}, {Malatesta}, {Peraza}, {Keil},
  {Silverberg}, {Jansen}, {Scheffler}, {Berdis}, {Larsen}, \&
  {Hilton}}]{davenport2014b}
{Davenport}, J.~R.~A., {et~al.} 2014, \apj, Accepted, arXiv \# 1411.3723

\bibitem[{{Delfosse} {et~al.}(2000){Delfosse}, {Forveille}, {S{\'e}gransan},
  {Beuzit}, {Udry}, {Perrier}, \& {Mayor}}]{delfosse2000}
{Delfosse}, X., {Forveille}, T., {S{\'e}gransan}, D., {Beuzit}, J.-L., {Udry},
  S., {Perrier}, C., \& {Mayor}, M. 2000, \aap, 364, 217

\bibitem[{{Dobler} {et~al.}(2006){Dobler}, {Stix}, \&
  {Brandenburg}}]{dobler2006}
{Dobler}, W., {Stix}, M., \& {Brandenburg}, A. 2006, \apj, 638, 336

\bibitem[{{Donati} \& {Brown}(1997)}]{donati_brown1997}
{Donati}, J.-F., \& {Brown}, S.~F. 1997, \aap, 326, 1135

\bibitem[{{Donati} \& {Collier Cameron}(1997)}]{donati1997}
{Donati}, J.-F., \& {Collier Cameron}, A. 1997, \mnras, 291, 1

\bibitem[{{Foreman-Mackey} {et~al.}(2012){Foreman-Mackey}, {Hogg}, {Lang}, \&
  {Goodman}}]{emcee}
{Foreman-Mackey}, D., {Hogg}, D.~W., {Lang}, D., \& {Goodman}, J. 2012, ArXiv
  e-prints

\bibitem[{{Frasca} {et~al.}(2011){Frasca}, {Fr{\"o}hlich}, {Bonanno},
  {Catanzaro}, {Biazzo}, \& {Molenda-{\.Z}akowicz}}]{frasca2011}
{Frasca}, A., {Fr{\"o}hlich}, H.-E., {Bonanno}, A., {Catanzaro}, G., {Biazzo},
  K., \& {Molenda-{\.Z}akowicz}, J. 2011, \aap, 532, A81

\bibitem[{{Gastine} {et~al.}(2013){Gastine}, {Morin}, {Duarte}, {Reiners},
  {Christensen}, \& {Wicht}}]{gastine2013}
{Gastine}, T., {Morin}, J., {Duarte}, L., {Reiners}, A., {Christensen}, U.~R.,
  \& {Wicht}, J. 2013, \aap, 549, L5

\bibitem[{{Gizis} {et~al.}(2013){Gizis}, {Burgasser}, {Berger}, {Williams},
  {Vrba}, {Cruz}, \& {Metchev}}]{gizis2013}
{Gizis}, J.~E., {Burgasser}, A.~J., {Berger}, E., {Williams}, P.~K.~G., {Vrba},
  F.~J., {Cruz}, K.~L., \& {Metchev}, S. 2013, \apj, 779, 172

\bibitem[Hall \& Henry(1994)]{hall1994b} Hall, D.~S., \& Henry, G.~W.\ 1994, International Amateur-Professional Photoelectric Photometry Communications, 55, 51 

\bibitem[{{Hawley} {et~al.}(2014){Hawley}, {Davenport}, {Kowalski},
  {Wisniewski}, {Hebb}, {Deitrick}, \& {Hilton}}]{hawley2014}
{Hawley}, S.~L., {Davenport}, J.~R.~A., {Kowalski}, A.~F., {Wisniewski}, J.~P.,
  {Hebb}, L., {Deitrick}, R., \& {Hilton}, E.~J. 2014, \apj, Accepted, arXiv \#
  1410.7779

\bibitem[{{Henry} {et~al.}(1995){Henry}, {Eaton}, {Hamer}, \&
  {Hall}}]{henry1995}
{Henry}, G.~W., {Eaton}, J.~A., {Hamer}, J., \& {Hall}, D.~S. 1995, \apjs, 97,
  513

\bibitem[{{Hull} {et~al.}(2013){Hull}, {Plambeck}, {Bolatto}, {Bower},
  {Carpenter}, {Crutcher}, {Fiege}, {Franzmann}, {Hakobian}, {Heiles}, {Houde},
  {Hughes}, {Jameson}, {Kwon}, {Lamb}, {Looney}, {Matthews}, {Mundy}, {Pillai},
  {Pound}, {Stephens}, {Tobin}, {Vaillancourt}, {Volgenau}, \&
  {Wright}}]{hull2013}
{Hull}, C.~L.~H., {et~al.} 2013, \apj, 768, 159

\bibitem[{{Irwin} {et~al.}(2011){Irwin}, {Berta}, {Burke}, {Charbonneau},
  {Nutzman}, {West}, \& {Falco}}]{irwin2011}
{Irwin}, J., {Berta}, Z.~K., {Burke}, C.~J., {Charbonneau}, D., {Nutzman}, P.,
  {West}, A.~A., \& {Falco}, E.~E. 2011, \apj, 727, 56

\bibitem[{{Irwin} {et~al.}(2009){Irwin}, {Charbonneau}, {Nutzman}, \&
  {Falco}}]{irwin2009}
{Irwin}, J., {Charbonneau}, D., {Nutzman}, P., \& {Falco}, E. 2009, in IAU
  Symposium, Vol. 253, IAU Symposium, ed. F.~{Pont}, D.~{Sasselov}, \& M.~J.
  {Holman}, 37--43

\bibitem[{{Kipping}(2012)}]{kipping2012}
{Kipping}, D.~M. 2012, \mnras, 427, 2487

\bibitem[{{Kiraga} \& {Stepien}(2007)}]{kiraga2007}
{Kiraga}, M., \& {Stepien}, K. 2007, \actaa, 57, 149

\bibitem[{{Kitchatinov} \& {Olemskoy}(2011)}]{kitchatinov2011}
{Kitchatinov}, L.~L., \& {Olemskoy}, S.~V. 2011, \mnras, 411, 1059

\bibitem[{{K{\"u}ker} \& {R{\"u}diger}(2005)}]{kuker_rudiger2005}
{K{\"u}ker}, M., \& {R{\"u}diger}, G. 2005, Astronomische Nachrichten, 326, 265

\bibitem[{{K{\"u}ker} \& {R{\"u}diger}(2008)}]{kuker_rudiger2008}
---. 2008, Journal of Physics Conference Series, 118, 012029

\bibitem[{{K{\"u}ker} \& {R{\"u}diger}(2011)}]{kuker_rudiger2011}
---. 2011, Astronomische Nachrichten, 332, 933

\bibitem[{{L{\'e}pine} \& {Shara}(2005)}]{lepine2005}
{L{\'e}pine}, S., \& {Shara}, M.~M. 2005, \aj, 129, 1483

\bibitem[{Litzkow {et~al.}(1988)Litzkow, Livny, \& Mutka}]{condor-hunter}
Litzkow, M., Livny, M., \& Mutka, M. 1988, in Proceedings of the 8th
  International Conference of Distributed Computing Systems

\bibitem[{{Lomb}(1976)}]{lomb1976}
{Lomb}, N.~R. 1976, \apss, 39, 447

\bibitem[{{Lurie} {et~al.}(2015){Lurie}, {Davenport}, {Hawley}, {Wilkinson},
  {Wisniewski}, {Kowalski}, \& {Hebb}}]{lurie2015}
{Lurie}, J.~C., {Davenport}, J.~R.~A., {Hawley}, S.~L., {Wilkinson}, T.~D.,
  {Wisniewski}, J.~P., {Kowalski}, A.~F., \& {Hebb}, L. 2015, ArXiv e-prints

\bibitem[{{McQuillan} {et~al.}(2013){McQuillan}, {Aigrain}, \&
  {Mazeh}}]{mcquillan2013}
{McQuillan}, A., {Aigrain}, S., \& {Mazeh}, T. 2013, \mnras, 432, 1203

\bibitem[{{McQuillan} {et~al.}(2014){McQuillan}, {Mazeh}, \&
  {Aigrain}}]{mcquillan2014}
{McQuillan}, A., {Mazeh}, T., \& {Aigrain}, S. 2014, \apjs, 211, 24

\bibitem[{{Morin} {et~al.}(2008{\natexlab{a}}){Morin}, {Donati}, {Forveille},
  {Delfosse}, {Dobler}, {Petit}, {Jardine}, {Collier Cameron}, {Albert},
  {Manset}, {Dintrans}, {Chabrier}, \& {Valenti}}]{morin2008a}
{Morin}, J., {et~al.} 2008{\natexlab{a}}, \mnras, 384, 77

\bibitem[{{Morin} {et~al.}(2008{\natexlab{b}}){Morin}, {Donati}, {Petit},
  {Delfosse}, {Forveille}, {Albert}, {Auri{\`e}re}, {Cabanac}, {Dintrans},
  {Fares}, {Gastine}, {Jardine}, {Ligni{\`e}res}, {Paletou}, {Ramirez Velez},
  \& {Th{\'e}ado}}]{morin2008b}
---. 2008{\natexlab{b}}, \mnras, 390, 567

\bibitem[{{Morin} {et~al.}(2010){Morin}, {Donati}, {Petit}, {Delfosse},
  {Forveille}, \& {Jardine}}]{morin2010}
{Morin}, J., {Donati}, J.-F., {Petit}, P., {Delfosse}, X., {Forveille}, T., \&
  {Jardine}, M.~M. 2010, \mnras, 407, 2269

\bibitem[{{N.~Reid \& S.~L.~Hawley}(2000)}]{nlds}
{N.~Reid \& S.~L.~Hawley}, ed. 2000, {New light on dark stars : red dwarfs, low
  mass stars, brown dwarfs}

\bibitem[{{Nutzman} \& {Charbonneau}(2008)}]{nutzman2008}
{Nutzman}, P., \& {Charbonneau}, D. 2008, \pasp, 120, 317

\bibitem[{{Parker}(1955)}]{parker1955}
{Parker}, E.~N. 1955, \apj, 122, 293

\bibitem[{Quintana {et~al.}(2014)Quintana, Barclay, Raymond, Rowe, Bolmont,
  Caldwell, Howell, Kane, Huber, Crepp, Lissauer, Ciardi, Coughlin, Everett,
  Henze, Horch, Isaacson, Ford, Adams, Still, Hunter, Quarles, \&
  Selsis}]{kepler186f}
Quintana, E.~V., {et~al.} 2014, Science, 344, 277

\bibitem[{{Ramsay} {et~al.}(2013){Ramsay}, {Doyle}, {Hakala}, {Garcia-Alvarez},
  {Brooks}, {Barclay}, \& {Still}}]{ramsay2013}
{Ramsay}, G., {Doyle}, J.~G., {Hakala}, P., {Garcia-Alvarez}, D., {Brooks}, A.,
  {Barclay}, T., \& {Still}, M. 2013, \mnras, 434, 2451

\bibitem[Reiners \& Schmitt(2003)]{reiners2003} 
Reiners, A., \& Schmitt, J.~H.~M.~M.\ 2003, \aap, 398, 647 

\bibitem[{{Reiners}(2006)}]{reiners2006}
{Reiners}, A. 2006, \aap, 446, 267

\bibitem[{{Reinhold} \& {Reiners}(2013)}]{reinhold2013a}
{Reinhold}, T., \& {Reiners}, A. 2013, \aap, 557, A11

\bibitem[{{Reinhold} {et~al.}(2013){Reinhold}, {Reiners}, \&
  {Basri}}]{reinhold2013}
{Reinhold}, T., {Reiners}, A., \& {Basri}, G. 2013, \aap, 560, A4

\bibitem[{{Roettenbacher} {et~al.}(2013){Roettenbacher}, {Monnier}, {Harmon},
  {Barclay}, \& {Still}}]{roettenbacher2013}
{Roettenbacher}, R.~M., {Monnier}, J.~D., {Harmon}, R.~O., {Barclay}, T., \&
  {Still}, M. 2013, \apj, 767, 60

\bibitem[{{Sanchis-Ojeda} {et~al.}(2013){Sanchis-Ojeda}, {Winn}, {Marcy},
  {Howard}, {Isaacson}, {Johnson}, {Torres}, {Albrecht}, {Campante}, {Chaplin},
  {Davies}, {Lund}, {Carter}, {Dawson}, {Buchhave}, {Everett}, {Fischer},
  {Geary}, {Gilliland}, {Horch}, {Howell}, \& {Latham}}]{sanchis-ojeda2013}
{Sanchis-Ojeda}, R., {et~al.} 2013, \apj, 775, 54

\bibitem[{{Savanov} \& {Dmitrienko}(2011)}]{savanov2011}
{Savanov}, I.~S., \& {Dmitrienko}, E.~S. 2011, Astronomy Reports, 55, 890

\bibitem[{{Scargle}(1982)}]{scargle1982}
{Scargle}, J.~D. 1982, \apj, 263, 835

\bibitem[{{Schrijver} \& {Zwaan}(2000)}]{schrijver_zwaan2000}
{Schrijver}, C.~J., \& {Zwaan}, C. 2000, {Solar and Stellar Magnetic Activity}
  (New York : Cambridge University Press, 2000 (Cambridge astrophysics series ;
  34))

\bibitem[{{Semel}(1989)}]{semel1989}
{Semel}, M. 1989, \aap, 225, 456

\bibitem[{{Smith} {et~al.}(2012){Smith}, {Stumpe}, {Van Cleve}, {Jenkins},
  {Barclay}, {Fanelli}, {Girouard}, {Kolodziejczak}, {McCauliff}, {Morris}, \&
  {Twicken}}]{smith2012}
{Smith}, J.~C., {et~al.} 2012, \pasp, 124, 1000

\bibitem[{Thain {et~al.}(2005)Thain, Tannenbaum, \& Livny}]{condor-practice}
Thain, D., Tannenbaum, T., \& Livny, M. 2005, Concurrency - Practice and
  Experience, 17, 323

\bibitem[{{Walkowicz} {et~al.}(2011){Walkowicz}, {Basri}, {Batalha},
  {Gilliland}, {Jenkins}, {Borucki}, {Koch}, {Caldwell}, {Dupree}, {Latham},
  {Meibom}, {Howell}, {Brown}, \& {Bryson}}]{walkowicz2011}
{Walkowicz}, L.~M., {et~al.} 2011, \aj, 141, 50

\bibitem[{{West} {et~al.}(2008){West}, {Hawley}, {Bochanski}, {Covey}, {Reid},
  {Dhital}, {Hilton}, \& {Masuda}}]{west2008}
{West}, A.~A., {Hawley}, S.~L., {Bochanski}, J.~J., {Covey}, K.~R., {Reid},
  I.~N., {Dhital}, S., {Hilton}, E.~J., \& {Masuda}, M. 2008, \aj, 135, 785

\bibitem[{{Yadav} {et~al.}(2015){Yadav}, {Gastine}, {Christensen}, \&
  {Reiners}}]{yadav2015}
{Yadav}, R.~K., {Gastine}, T., {Christensen}, U.~R., \& {Reiners}, A. 2015,
  \aap, 573, A68

\bibitem[{{Zacharias} {et~al.}(2013){Zacharias}, {Finch}, {Girard}, {Henden},
  {Bartlett}, {Monet}, \& {Zacharias}}]{ucac4}
{Zacharias}, N., {Finch}, C.~T., {Girard}, T.~M., {Henden}, A., {Bartlett},
  J.~L., {Monet}, D.~G., \& {Zacharias}, M.~I. 2013, \aj, 145, 44

\end{thebibliography}
\end{document}